\documentclass[10pt,twocolumn,letterpaper]{article}

\usepackage{tmp}
\usepackage{times}
\usepackage{graphicx}
\usepackage{amsmath}
\usepackage{amssymb}
\usepackage{booktabs}
\usepackage{nicefrac}
\usepackage{pifont}
\usepackage[ruled]{algorithm2e}
\usepackage{multirow}
\usepackage{float}
\usepackage{cite}
\usepackage[pagebackref=true,breaklinks=true,letterpaper=true,colorlinks,bookmarks=false]{hyperref}

\tmpfinalcopy 

\iftmpfinal\fi

\newcommand{\mat}[1]{\boldsymbol{#1}}
\begin{document}

\title{MANAS: Multi-Scale and Multi-Level Neural Architecture Search for Low-Dose CT Denoising}

\author{Zexin Lu$^1$, Wenjun Xia$^1$, Yongqiang Huang$^1$, Hongming Shan$^2$, Hu Chen$^1$, Jiliu Zhou$^1$, Yi Zhang$^1$\\
$^1$College of Computer Science, Sichuan University, Chengdu, China \\
$^2$Institute of Science and Technology for Brain-inspired Intelligence, Fudan University, China\\
{\tt\small zexinlu.scu@gmail.com, hmshan@fudan.edu.cn, yzhang@scu.edu.cn}
}

\maketitle
\iftmpfinal\thispagestyle{empty}\fi

\begin{abstract}
    Lowering the radiation dose in computed tomography (CT) can greatly reduce the potential risk to public health. However, the reconstructed images from the dose-reduced CT or low-dose CT (LDCT) suffer from severe noise, compromising the subsequent diagnosis and analysis. Recently, convolutional neural networks have achieved promising results in removing noise from LDCT images; the network architectures used are either handcrafted or built on top of conventional networks such as ResNet and U-Net. Recent advance on neural network architecture search (NAS) has proved that the network architecture has a dramatic effect on the model performance, which indicates that current network architectures for LDCT may be sub-optimal. Therefore, in this paper, we make the first attempt to apply NAS to LDCT and propose a multi-scale and multi-level NAS for LDCT denoising, termed MANAS. On the one hand, the proposed MANAS fuses features extracted by different scale cells to capture multi-scale image structural details. On the other hand, the proposed MANAS can search a hybrid cell- and network-level structure for better performance.  
    Extensively experimental results on three different dose levels demonstrate that the proposed MANAS can achieve better performance in terms of preserving image structural details than several state-of-the-art methods. In addition, we also validate the effectiveness of the multi-scale and multi-level architecture for LDCT denoising.
\end{abstract}

\section{Introduction}

Nowadays, X-ray computed tomography (CT) has been widely used in medical field. Since the radiation generated by CT scanning may cause irreversible damage to human body, more and more researchers pay attention to low-dose CT (LDCT) under the well-known as low as reasonably achievable or ALARA principle~\cite{brenner2007computed}. The most common way to lower the radiation dose is to reduce the X-ray flux by decreasing the operating current. As the radiation dose decreases, the imaging quality is contaminated by severe noise and artifacts, which compromise the subsequent clinical diagnoses. \emph{How to achieve a pleasing image quality in LDCT remains active and challenging. }

To solve this problem, numerous algorithms have been developed over the last decade. These methods can be generally divided into three categories: 1) sinogram domain filtration, 2) iterative reconstruction, and 3) image post-processing.
Sinogram domain filtration directly processes either raw data or log-transformed data with a specific filter and then performs image reconstruction using filtered back projection (FBP). Typical methods include structural filtering~\cite{balda2012ray}, bilateral filtering~\cite{manduca2009projection}, and penalized weighted least-squares~\cite{wang2006penalized}. However, these kinds of methods usually suffer from spatial resolution loss once the edges in the sinogram are smoothed.
Iterative reconstruction is effective at suppressing the noise and artifacts. As a representative method for LDCT, the model-based iterative reconstruction method repeatedly performs forward- and back-projection and computes the regularizers in image domain~\cite{chen2018learn,xia2019spectral,chen2020airnet}, as a result, the computational burden of iterative reconstruction methods is tremendous, hampering its wide application in clinical use. In addition, these methods often suffer from the difficulty of accessing raw data since the details of the scanner geometry and data correction are usually not available to users. Significantly different from the above two categories, the image post-processing method is an efficient alternative, which does not rely on raw data and can be easily integrated into the current CT pipeline. Inspired by the idea of spare representation, several classic image denoising algorithms, such as non-local means, dictionary learning, and block-matching 3D (BM3D), which have attained a state-of-the-art performance~\cite{feruglio2010block,sheng2014denoised,kang2013image,buades2005non}. However, different from the assumption of traditional natural image denoising, the noise and artifacts in LDCT images do not obey any statistical distribution and cannot be accurately determined, which makes these image denoising methods have certain limitations.

Recently, deep learning (DL) has received increasingly attention in many fields of computer vision, including object detection~\cite{ren2015faster,redmon2016you,liu2016ssd}, semantic segmentation~\cite{ronneberger2015u} and image restoration~\cite{zhang2017beyond,wang2018esrgan,ulyanov2018deep}. In the field of LDCT denoising, a growing number of DL-based methods have been proposed~\cite{wang2018image,wang2019machine,wang2020deep}. Since the DL-based post-processing methods are computationally efficient and do not need any assumption on the noise distribution, it has become a hot research topic. Such methods usually use a classic or modified end-to-end network architecture and take the paired data for supervised learning. Typical network 
architectures include VGG~\cite{simonyan2014very}, AlexNet~\cite{krizhevsky2012imagenet}, ResNet~\cite{he2016deep}, DenseNet~\cite{huang2017densely} and U-Net~\cite{ronneberger2015u}. However, there are no further studies on the design of network architecture and its corresponding influence on denoising performance. All these networks are handcrafted, which are limited by the researcher’s experience and computing resources. Manually designed networks usually face two main problems. First, due to the differences in target datasets, substantial efforts are required to select the appropriate network structure according to the characteristics of the data. Second, it is also difficult to balance the tradeoff between the scale and performance of the network. 
Compared with the manually-designed methods, neural architecture search (NAS), which does not rely on expert experience and knowledge, has attracted significant interest and achieved competitive performance in the fields of image classification~\cite{liu2018darts,liang2019darts+,cai2018proxylessnas} and segmentation~\cite{li2020learning,liu2019auto,weng2019unet,yan2020ms}. Current NAS methods can be roughly classified into reinforcement learning~\cite{baker2016designing,zhong2018practical,zoph2016neural,zoph2018learning}, evolutionary algorithms~\cite{real2017large,xie2017genetic,miikkulainen2019evolving}, or gradient-based methods~\cite{liang2019darts+,liu2018darts,yan2020ms,weng2019unet}. The first two kinds are computational expensive and possibly not suitable for pixel-level denoising task. In this paper, inspired by the idea of~\cite{liu2018darts,liu2019auto,yan2020ms}, we develop a nested-like super-net that contains all candidate operations as shown in Fig.~\ref{fig:architecture} to enlarge the search space and apply the continuous relaxation method~\cite{liu2018darts} on both inner cell and outer network levels. Once the searching stage finishes, the Dijkstra algorithm is adopted to find the optimal sub-net with corresponding cells. 

Our contributions are summarized as follows:
\begin{itemize}
    \setlength{\itemsep}{0pt}
    \setlength{\parsep}{0pt}
    \setlength{\parskip}{0pt}
    \item We propose a multi-scale and multi-level NAS, termed MANAS, for low-dose CT denoising. To the best of our knowledge, this is the first attempt to extend NAS to LDCT denoising.
    \item We propose a multi-scale fusion cell to leverage the features extracted from different scales in the searching stage.
    \item We propose a multi-level super-net integrating both cell- and network-level search to enlarge the search space. It enables the algorithm to easily find a more efficient sub-net from the super-net defined in Fig.~\ref{fig:architecture} while handling data from different dose levels. The training of the proposed MANAS can be done in 2 RTX 8000 GPU days under the continuous relaxation method.
    \item Extensive experiments demonstrate that the network architectures searched by our method provide better visual effects and reduce the scale of the network compared with several state-of-the-art models.
\end{itemize}
The rest of this paper is organized as follows. In Sec.~\ref{sec:relatedwork}, recent advances in DL-based LDCT denoising and NAS are reviewed. Then we elaborate on the specific implementation of the proposed MANAS in Sec.~\ref{sec:method}. Sec.~\ref{sec:experiment} presents the experimental design and representative results are presented, which is followed by a concluding summary in Sec.~\ref{sec:conclusion}.
\section{Related Works}\label{sec:relatedwork}

\subsection{Deep Learning based LDCT Image Denoising}

Recently, thanks to the rapid development of convolution neural networks (CNNs), the performance of LDCT image denoising algorithms have been significantly improved. As the first work, Chen~\etal introduced the famous super-resolution CNN or SRCNN~\cite{dong2014learning} into LDCT restoration and then proposed a residual encoder-decoder CNN or RED-CNN for LDCT with encouraging results.  Kang~\etal~\cite{kang2018deep} introduced wavelet transform to U-Net~\cite{weng2019unet}. With the emergence of generative adversarial networks (GANs), Yang~\etal~proposed to combine Wasserstein GANs and perceptual loss, termed WGAN-VGG, which can recover the mottle-like texture in CT images~\cite{yang2018low}. Shan~\etal extended this model for 3D LDCT with an efficient training strategy based on transfer learning~\cite{shan20183}, and then proposed a modularized adaptive processing neural network (MAP-NN), which performs an end-to-process mapping with a modularized neural network and allows radiologists in the loop to optimize the denoising depth in a task-specific fashion~\cite{shan2019competitive}. Inspired by the successful applications in computer vision, attention mechanism was introduced into LDCT to retrieve pixels with strong relationships across long distance and achieved promising results~\cite{li2020sacnn, huang2020cagan}. In addition, CycleGAN was also used to mitigate the difficulty in acquiring paired training samples~\cite{kang2019cycle,gu2020adain}. Compared with these handcrafted network architectures, our work does not rely on expert experience and can get better LDCT denoising performance.
\begin{figure*}[t]
   \begin{center}
   \includegraphics[width=0.83\linewidth]{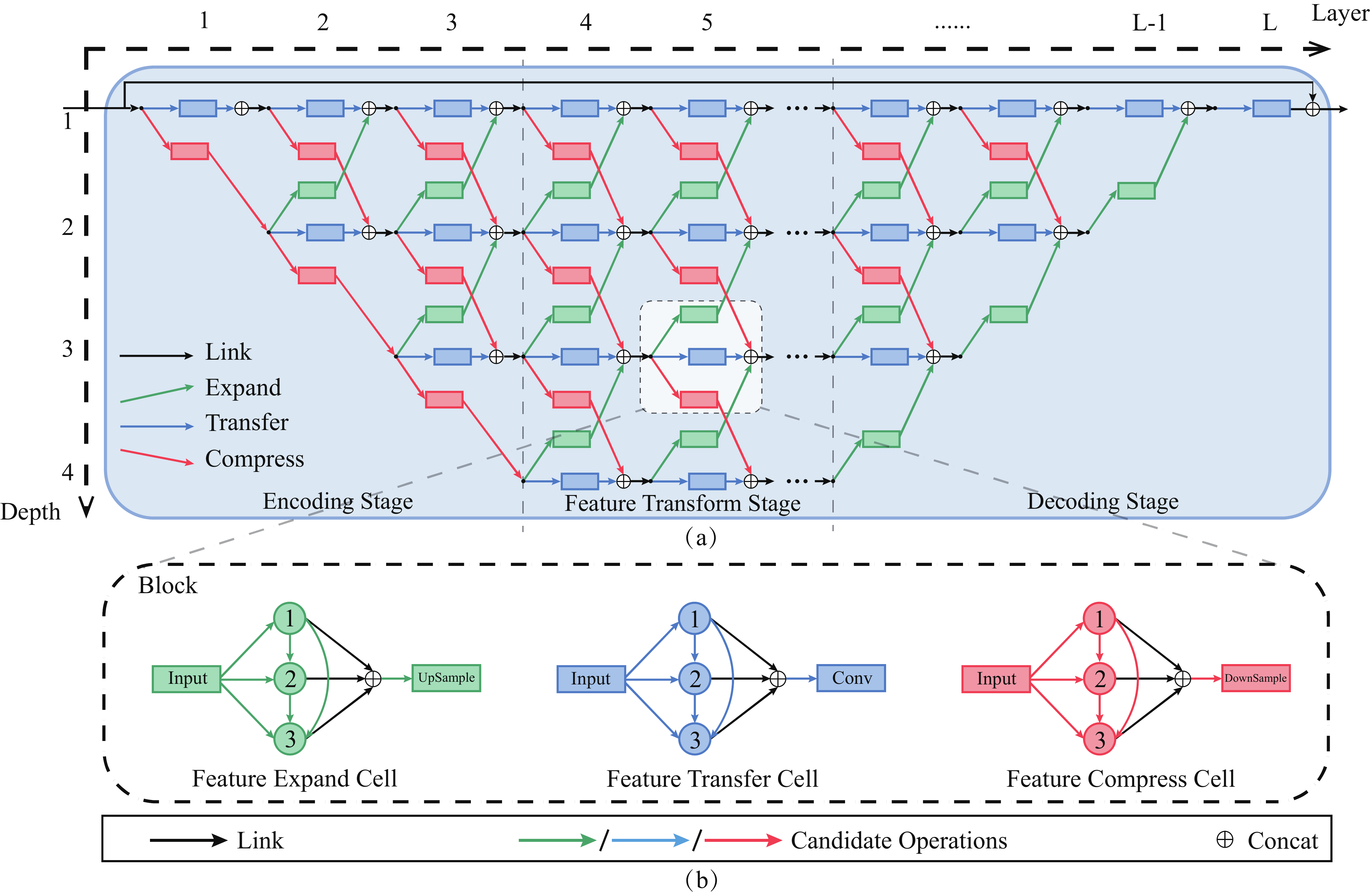}
   \end{center}
      \caption{ The overall framework of the proposed MANAS. (a) The architecture of the super-net and (b) the basic block involving three kinds of cells.}
   \label{fig:architecture}
   \end{figure*}
\subsection{Network Architecture Search (NAS)}
NAS is dedicated to automating the design of neural network architectures, so that this tedious work, which heavily depends on researchers' practical experience, can be significantly alleviated. Several studies proposed to optimize the network architectures using the basic operations in the evolutionary algorithms~\cite{real2017large,xie2017genetic}. Reinforcement learning methods, e.g., Q-Learning~\cite{zhong2018practical,baker2016designing} and policy gradients~\cite{zoph2016neural,zoph2018learning}, aim to train a recurrent neural network as a controller to generate specific architectures. Different from the above two strategies that are computationally intensive and time-consuming in the search stage, differentiable architecture search (DARTS) employs continuous relaxation to perform an efficient search of the cell architecture using gradient descent, which makes it possible to train NAS networks on a single GPU.
Based on the fundamental DARTS, many efforts were made for various tasks. For example, in~\cite{liu2019auto}, Liu~\etal proposed the Auto-DeepLab to search in a hierarchical architecture search space for semantic image segmentation. Ghiasi~\etal proposed NAS-FPN to learn scalable feature pyramid architecture for object detection~\cite{ghiasi2019fpn}. In~\cite{zhang2020memory}, HiNAS is proposed for natural image denoising. 
In the field of medical imaging, most works were proposed for image segmentation and classification~\cite{yan2020ms,weng2019unet,yu2020c2fnas,liu2020scam,dong2019neural,guo2020organ}. Only two works most relevant to our work are dedicated to MRI reconstruction using DARTS~\cite{huang2020enhanced,yan2020neural}. Nevertheless, both methods adopt a simple plain network architecture combined with residual block, which only searches the repeatable cell structure and ignores the impact of the outer network-level structure. Furthermore, as demonstrated by~\cite{chen2017low, huang2020cagan}, fusing the features extracted from different scales is helpful to recover more details. At the same time, when manually designing the network, people who lack practical experience will cause the designed structure to have a certain degree of redundancy and increase the number of additional network parameters. Despite these shortcomings mentioned above, these works have shown encouraging potential in the field of medical image reconstruction. In this study, motivated by these pioneering works~\cite{yan2020ms,liu2019auto,liu2018darts}, we make the first attempt to integrate NAS for LDCT denoising aided by both inner cell- and outer network-levels search and multi-scale feature fusion. Our work follows the formulation of differentiable NAS methods. Compared with other NAS models, our work is to jointly search the cells and network architecture. We also construct a nested-like super-net to enlarge the search space to make the model capture more features from different scales.


\section{Method}\label{sec:method}

\subsection{Problem Formulation}

Assuming that $\mat{I}_{\mathrm{LD}} \in \mathbb{R}^{m \times n}$ is an LDCT image of size $m\times n$ and $\mat{I}_{\mathrm{ND}} \in \mathbb{R}^{m \times n}$ is the corresponding normal-dose CT (NDCT) image, the LDCT image restoration is formulated to find a function $f$, which maps an LDCT image to its normal-dose counterpart:
\begin{align}
\mathop{\arg\min}_{f} \mathcal{L}_\mathrm{MSE}(\mat{I}_{\mathrm{LD}}, \mat{I}_{\mathrm{ND}})= \| f(\mat{I}_{\mathrm{LD}})-\mat{I}_{\mathrm{ND}} \|_2^2, \label{eq:objective_function}
\end{align}
where $\mathcal{L}_\mathrm{MSE}$ represents the widely-used mean squared error (MSE).

Instead of finding an explicit function $f$, we use a convolution neural network to automatically achieve this process in a data-driven fashion. Unlike the traditionally manual design of the neural network, we use the NAS method to find a suitable network. That is, we try to find $f$ using an automatic method.

\subsection{The Proposed MANAS}

Similar to~\cite{liu2018darts,liu2019auto}, in our MANAS, the gradient-based strategy is adopted to search for the basic cells' architecture and the sub-net architecture from the nested-liked super-net. Current researches provide limited choices of network architecture for low-level tasks, especially for medical images~\cite{huang2020enhanced,zhang2020memory}. Inspired by existing DL-based LDCT image restoration, multi-scale transformation can extract features from different spatial resolutions and is useful to recover more details in multiple scales. Following the idea in~\cite{liu2019auto,yan2020ms}, a more flexible search space is proposed for LDCT image restoration. A nested-liked super-net is built in Fig.~\ref{fig:architecture}(a), which can be divided into three stages: 1) encoding stage, 2) feature transformation stage, and 3) decoding stage. In the encoding stage, the details extracted from different scales are encoded using various candidate operations. In the feature transformation stage, the multi-scale features are extracted and filtered to suppress the noise and artifacts. In the decoding stage, the features processed from the previous stage are decoded to recover the details with the best decoding strategy using the candidate operations. In each stage, a multi-scale fusion block is adopted to make full use of the multi-scale features.

\begin{figure}[t]
   \begin{center}
      \includegraphics[width=1.0\linewidth]{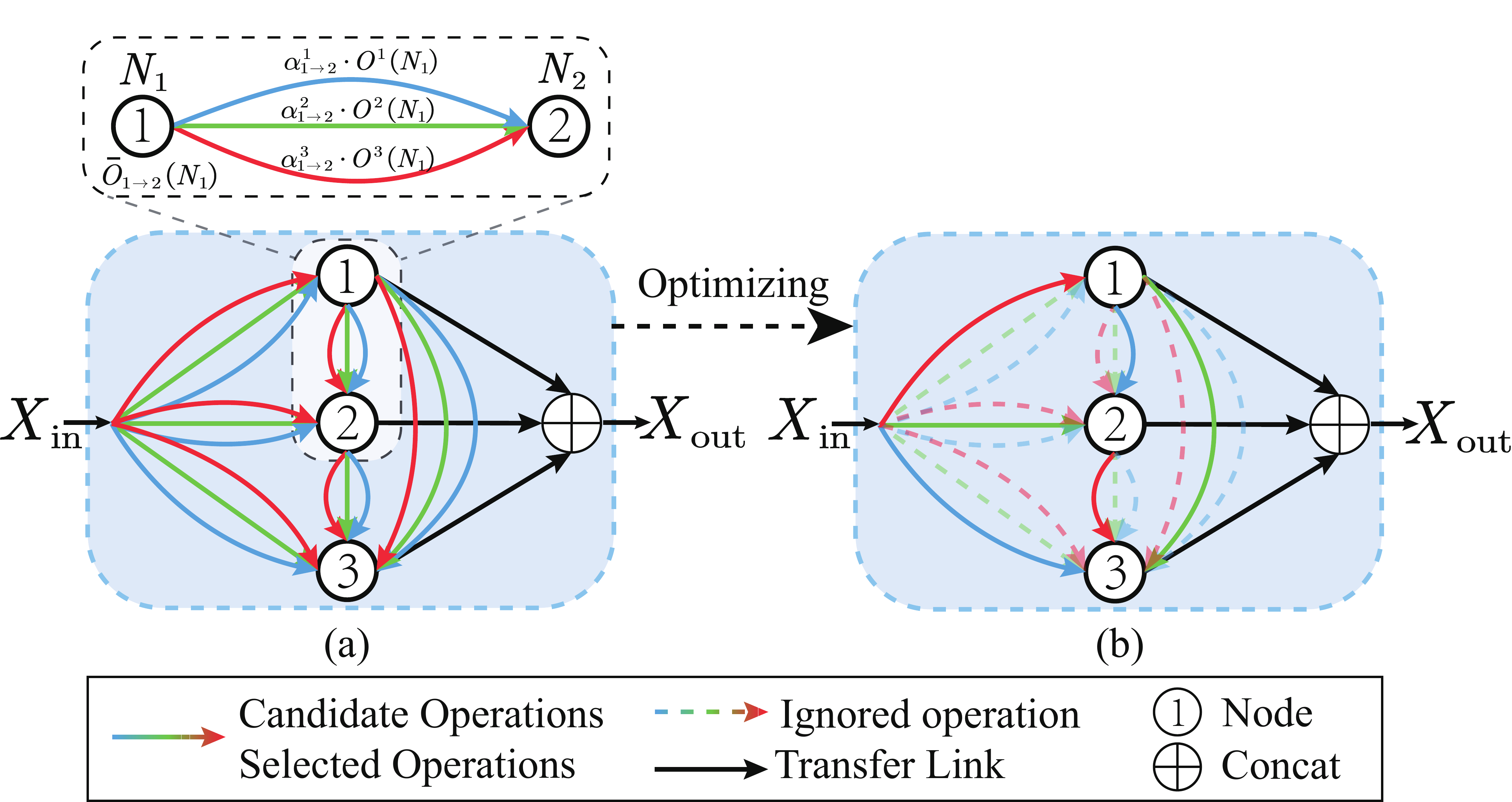}
   \end{center}
      \caption{Inner cell construct stage and search stage.}
   \label{fig:innercell}
\end{figure}

This subsection first introduces how to search basic cells using continuous relaxation, and then shows how to construct a basic block from cells, and finally describes how to form a multi-level super-net.

\subsubsection{Basic Cell}
We define three different types of cells for multi-scale image feature extraction. Following~\cite{liu2018darts}, we employ the continuous relaxation strategy to search for the appropriate cells. Typically, a super-cell is represented by a directed acyclic graph (DAG) with $P$ nodes. Fig.~\ref{fig:innercell}(a) and (b) show a super-cell containing all candidate operations and one example of the optimized result, respectively. 
For simplicity, we only show three nodes in Fig.~\ref{fig:innercell}. Our target is to map the input $\mat{X}_\mathrm{in}$ to the output $\mat{X}_\mathrm{out}$ through the cell. The output of one cell is represented as follows:
\begin{align}
\mat{X}_\mathrm{out}=\mathrm{Concat}(\mat{N}_1, \mat{N}_2, \cdots, \mat{N}_P), \label{eq:output_c}
\end{align}
where $\mat{N}_i$ denotes the output of $i$-th node and is defined as $\mat{N}_i=\sum_{\mat{N}_j{\in}\Omega_i} \bar{O}_{j{\to}i}(\mat{N}_j), i=1,2,\cdots,P$, $j<i$. $\Omega_i$ is the set of all the nodes before node $i$. $\bar{O}_{j{\to}i}$ is the set of all the candidate operations from node $j$ to node $i$ as
\begin{align}
\bar{O}_{j{\to}i}=\sum_{t=1}^{T} \alpha_{j{\to}i}^tO^t(\mat{N}_j), \label{eq:condidate_op}
\end{align}
where $\{O^1, O^2, \cdots, O^T\}$ represents a set of $T$ candidate operations and $\alpha_{j{\to}i}^k$ denotes the weight for the corresponding candidate operations.

For simplicity, we define $\mat{N}_0=\mat{X}_\mathrm{in}$, when $i>0, \Omega_i=\{\mat{N}_0, \mat{N}_1, \cdots, \mat{N}_{i-1}\},$ which means that first node only receives $\mat{X}_\mathrm{in}$ as input, and the other nodes receive all the preceding tensors (including $\mat{X}_\mathrm{in}$) as input.

According to~\cite{liu2018darts}, the softmax function is applied after all possible operations to make the search space continuous:
$
\mathrm{Softmax}(\alpha_{j{\to}i}^t)=\frac{\exp(\alpha_{j{\to}i}^t)}{\sum_{t-1}^T \exp(\alpha_{j{\to}i}^t)}. 
$

\begin{table}[t]
    \centering
    \begin{tabular}{r|l}
     \textbf{Acronym} & \textbf{Meaning} \\ \hline
     C@3 & $3\times3$ convolution \\ 
     SC@3 & $3\times3$ separable convolution\\ 
     SC@5 & $5\times5$ separable convolution\\ 
     DC@3 & $3\times3$  convolution with dilation rate 2\\ 
     DC@5 & $5\times5$  convolution with dilation rate 2\\ 
     Skip & Skip connection\\ 
     None & Without any connection and return zero\\ 
    \end{tabular}
    \caption{Candidate operations}
    \label{tab:candidate_operations}
\end{table}

Based on the recent works on DL-based LDCT image restoration, several typical candidate operations are shown in Table~\ref{tab:candidate_operations}.

Each operation starts with a ReLU layer and a $1\times1$ Conv layer is followed to keep the number of the features consistent. Batch normalization is excluded since it does not perform well on the PSNR oriented tasks~\cite{lim2017enhanced}.

\subsubsection{Multi-Scale Block}

To better use the features extracted from different kinds of cells, we construct a multi-scale block. In different scales, the model can capture different image features, which will help the model to recover better image details~\cite{bao2020real}. But not all features from different scales contribute a lot in one block, so we add architecture weights $\beta$ to select the most contributing cell. As illustrated in Fig.~\ref{fig:architecture}(b), the basic block in super-net is defined as:
\begin{align}
\mat{Z}_d^l&=\beta_{(d-1){\to}d}^l f_\mathrm{e}(\mat{Z}_{d-1}^{l-1};\alpha_\mathrm{e}) + \beta_{d \to d}^l f_\mathrm{t}(\mat{Z}_d^{l-1};\alpha_\mathrm{t}) \notag\\ 
&+ \beta_{(d+1) \to d}^l f_\mathrm{c}(\mat{Z}_{d+1}^{l-1}; \alpha_\mathrm{c}),   
\label{eq:super_net}
\end{align}
where $\mat{Z}_d^l$ is the output for the cell located in layer $l$ and depth $d$. $f_\mathrm{e}$, $f_\mathrm{t}$, and $f_\mathrm{c}$ denote the feature expand cell, feature transfer cell and feature compress cell, respectively. $(d+1) \to d$ means from depth $d+1$ to $d$ where $d \in \{1,2,\cdots,D\}$ and $D$ is the depth of the super-net. $\alpha$ and $\beta$ are the weight of the candidate operations in the cells and the weight for different cells in the block, respectively. It should be noted that not all the blocks have same cells as shown in the top and the bottom of the network. Meanwhile, the softmax function is applied to normalize $\beta$ as $\mathrm{Softmax}(\beta)=\frac{\exp(\beta)}{\sum_{\beta \in \mathcal{B}} \exp(\beta)}$, where $\mathcal{B}$ is the set of candidate cells in current block.

\subsubsection{Multi-Level Super-Net}

In our model, as the number of network layers increases by one, the number of features is doubled. In the same layer, as the network depth increases by one, the image size is reduced by half. Otherwise, the image size is doubled. By stacking the basic blocks into a nested-liked network, the multi-scale features extracted for the images can be fully exploited. Each block receives the fused features from the previous blocks in the super-net and then outputs processed features to the subsequent blocks. Aided by this architecture, the super-net can provide a more extensive search space covering different scales---from the network, block, and cell to operations, which makes the NAS algorithm easy to search for an efficient network architecture.

MANAS has the following multi-level feature:

\noindent\textbf{Cell-level.}\quad Instead of using three different cell structures for one  block, we only create three different kinds of inner cells to construct all the blocks in our model to save computational resources. Once optimized, the candidate operation with the largest $\alpha$ is chosen on each edge in the cell.

\noindent\textbf{Network-level.}\quad To optimize the super-net and determine the network architecture at the network level, we transform this super-net into a DAG, then we can construct all the paths to connect the input and final output. By calculating the accumulated $\beta$ using the Dijkstra algorithm, top-$K$ paths are selected as the best sub-net architecture. In this paper, $K$ is set to 5 throughout experiments. 

\subsection{Loss Function and Optimization}

\noindent\textbf{Loss function.}\quad  In our model, mean squared error and perceptual loss are used as the hybrid loss function, which is defined as:
\begin{align}
\mathcal{L}=\mathcal{L}_\mathrm{MSE}( f(\mat{I}_{\mathrm{LD}}), \mat{I}_{\mathrm{ND}}) +\lambda  \mathcal{L}_{\mathrm{PL}}(f(\mat{I}_{\mathrm{LD}}), \mat{I}_{\mathrm{ND}}), \label{eq:loss}
\end{align}
where the perceptual loss is defined as $\mathcal{L}_{\mathrm{PL}}(f(\mat{I}_{\mathrm{LD}}), \mat{I}_{\mathrm{ND}}) = \| \phi(f(\mat{I}_{\mathrm{LD}})) - \phi(\mat{I}_{\mathrm{ND}}) \|_2^2$ and $\phi$ is the pretrained VGG-16 network with parameters fixed. $\lambda$ is a weighting coefficient, which is empirically set to be $1 \times 10^{-4}$ in this paper.

\noindent\textbf{Optimization.}\quad Similar to other works~\cite{liu2019auto,yan2020ms}, the continuous relaxation strategy is adopted to optimize the cell and network parameters. It has been proven that this method can optimize these parameters efficiently using a gradient descent algorithm. Here we use the first-order approximation in DARTS and split the training set into train-data and arch-data; the ratio of train-data to arch-data is $7:1$. The loss function in Eq.~\eqref{eq:loss} is computed on the train-data to optimize the network parameters and on arch-data to optimize the architecture weights; these two processes are optimized alternatively.


\section{Experiment}\label{sec:experiment}

\subsection{Dataset}

The ``2016 NIH-AAPM-Mayo Clinic Low-Dose CT Grand Challenge''~\cite{mccollough2016tu} dataset is chosen to evaluate the performance of our proposed model. The dataset has 5,936 full-dose CT images from 10 patients. In our experiment, we choose 400 images from 8 patients randomly as the training set. Particularly, 50 images from the training set as the arch-data. 100 images from the remaining two patients as the testing set. The images are resized to $256 \times 256$. Poisson noise and electronic noise are added into the measured projection data to simulate the LDCT images with different dose levels:
$\hat{y}=\ln\frac{p}{\mathcal{P}(p\exp(-y))+\mathcal{N}(0,\sigma_e^2)}$,
where $\mathcal{P}$ and $\mathcal{N}$ represent the Poisson distribution and Gaussian distribution, respectively, 
$p$ is the number of photons before the X-ray penetrates the object, $\sigma_e^2$ is the variance of electronic noise generated from equipment measurement error, and $y$ represents the noise-free projection. The X-ray intensity of a normal dose is set to $p_0=1 \times 10^6$~\cite{niu2014sparse}. LDCT images with three different dose levels, 10\%, 5\% and 2.5\%, which corresponds to $p=0.1 \times p_0$, $p_0=0.05\times p_0$ and $p=0.025 \times p_0$, respectively, are simulated. In all experiments, we set $\sigma_e^2=10$.

\subsection{Implementation Details}
\begin{figure*}[!htb]
   \begin{center}
   \includegraphics[width=1\linewidth]{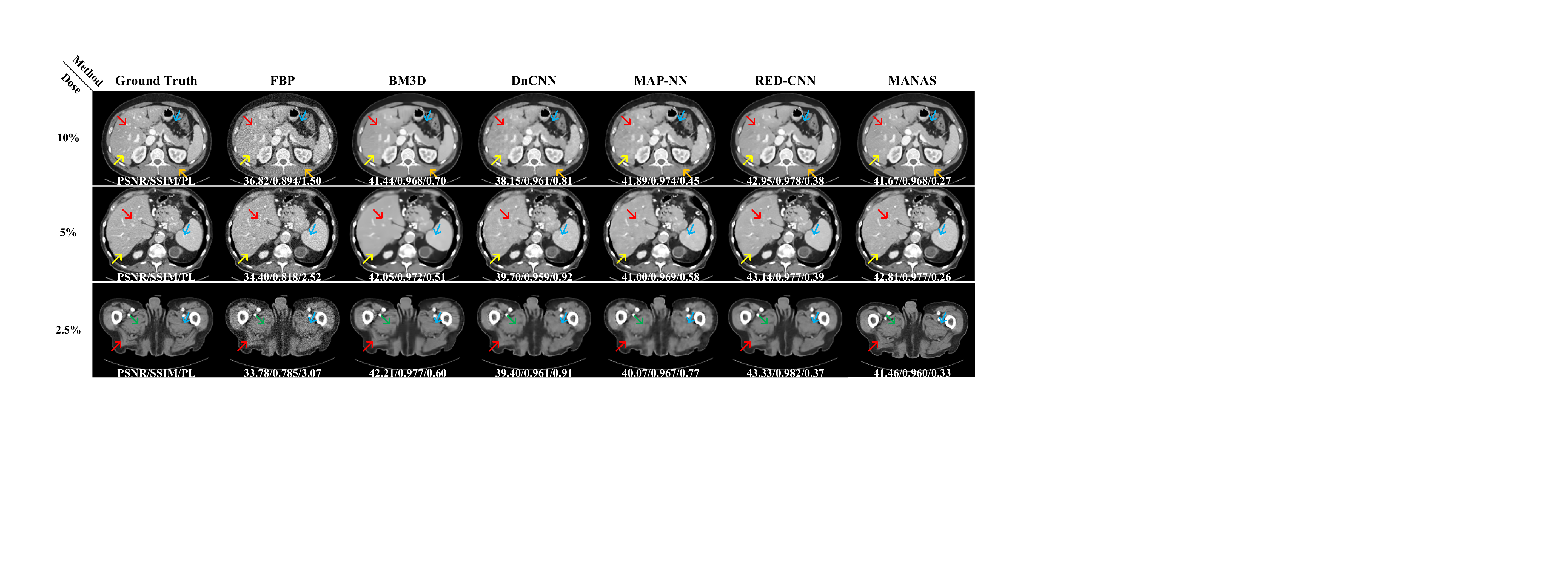}
   \end{center}
      \caption{Image denoising with 10\%, 5\%, and 2.5\% dose data by different methods.}
\label{fig:denoising_10to2.5}
\end{figure*}
\begin{table*}[!htb]
   \centering
    \resizebox{\textwidth}{!}{
      \begin{tabular}{rccccccccccc}
         \toprule
         \multirow{2}{*}{Dose} & \multicolumn{3}{c}{10\%} &  & \multicolumn{3}{c}{5\%} &  & \multicolumn{3}{c}{2.5\%} \\ 
         \cline{2-4} \cline{6-8} \cline{10-12}
          & PSNR $\uparrow$ & SSIM $\uparrow$ & PL $\downarrow$ &  & PSNR $\uparrow$ & SSIM $\uparrow$ & PL $\downarrow$ &  & PSNR $\uparrow$ & SSIM $\uparrow$ & PL $\downarrow$\\ \midrule
         FBP & 38.04{\footnotesize $\pm$0.68} & 0.908{\footnotesize $\pm$0.013} & 1.45{\footnotesize $\pm$0.18} &  & 35.25{\footnotesize $\pm$0.74} & 0.846{\footnotesize $\pm$0.021} & 2.32{\footnotesize $\pm$0.23} &  & 32.36{\footnotesize $\pm$0.77} & 0.756{\footnotesize $\pm$0.030} & 3.44{\footnotesize $\pm$0.27} \\
         BM3D\cite{dabov2007image} & 43.45{\footnotesize $\pm$0.55} & 0.980{\footnotesize $\pm$0.003} & 0.51{\footnotesize $\pm$0.09} &  & 42.24{\footnotesize $\pm$0.62} & 0.977{\footnotesize $\pm$0.004} & 0.50{\footnotesize $\pm$0.07} &  & 39.86{\footnotesize $\pm$0.94} & 0.960{\footnotesize $\pm$0.009} & 0.76{\footnotesize $\pm$0.12} \\
         DnCNN~\cite{zhang2017beyond} & 39.81{\footnotesize $\pm$0.76} & 0.970{\footnotesize $\pm$0.003} & 0.63{\footnotesize $\pm$0.09} &  & 39.21{\footnotesize $\pm$0.81} & 0.963{\footnotesize $\pm$0.004} & 0.84{\footnotesize $\pm$0.13} &  & 38.62{\footnotesize $\pm$0.66} & 0.957{\footnotesize $\pm$0.006} & 1.03{\footnotesize $\pm$0.16} \\
         RED-CNN~\cite{chen2017low} & \textbf{44.86{\footnotesize $\pm$0.60}} & \textbf{0.986{\footnotesize $\pm$0.003}} & 0.20{\footnotesize $\pm$0.05} &  & \textbf{43.40{\footnotesize $\pm$0.62}} & \textbf{0.982{\footnotesize $\pm$0.003}} & 0.31{\footnotesize $\pm$0.06} &  & \textbf{41.93{\footnotesize $\pm$0.64}} & \textbf{0.976{\footnotesize $\pm$0.004}} & 0.45{\footnotesize $\pm$0.09} \\
         MAP-NN~\cite{shan2019competitive} & 43.40{\footnotesize $\pm$0.50} & 0.983{\footnotesize $\pm$0.003} & 0.27{\footnotesize $\pm$0.05} &  & 41.15{\footnotesize $\pm$0.48} & 0.974{\footnotesize $\pm$0.004} & 0.47{\footnotesize $\pm$0.08} &  & 38.95{\footnotesize $\pm$0.45} & 0.959{\footnotesize $\pm$0.006} & 0.95{\footnotesize $\pm$0.13} \\
         MANAS(ours) & 43.22{\footnotesize $\pm$0.53} & 0.973{\footnotesize $\pm$0.003} &  \textbf{0.17{\footnotesize $\pm$0.03}} &  & 42.02{\footnotesize $\pm$0.62} & 0.981{\footnotesize $\pm$0.003} & \textbf{0.23{\footnotesize $\pm$0.04}} &  & 40.21{\footnotesize $\pm$0.52} & 0.955{\footnotesize $\pm$0.006} & \textbf{0.42{\footnotesize $\pm$0.07}} \\ \bottomrule
         \end{tabular}
  }
   \caption{Quantitative result (Mean$\pm$SD) of different methods on the testing sets of three low dose levels. The configuration of MANAS is $L=12$, $D=4$, $K=5$.}
   \label{tab:quantitative_methods}
\end{table*}

In this work, the number of nodes in each basic cell is set to 3. The numbers of layers $L$ and depths $D$ are set to 12 and 4, respectively. Top-$5$ paths are selected to form the best sub-net architecture.
The super-net was trained for 200 epochs with batch size 2. Two different Adam~\cite{kingma2014adam} optimizers $\mathrm{Optm}_\mathrm{n}$ and $\mathrm{Optm}_\mathrm{a}$ were used to optimize the network parameters in super-net and the architecture weights in the network searching stage, respectively. For $\mathrm{Optm}_\mathrm{n}$ the initial learning rate is set to $1 \times 10^{-4}$ and the decay to $1 \times 10^{-7}$ using the cosine annealing strategy~\cite{loshchilov2016sgdr}. For $\mathrm{Optm}_\mathrm{a}$ the initial learning rate is set to $1 \times 10^{-4}$ and the other configurations are set as default. 
Existing researches have shown that with the increase of training epochs, the network tends to select skip-connections in the cell search stage, which causes model collapse~\cite{chu2020fair,liang2019darts+}. To avoid model collapse, early stopping is employed to restrict the number of skip-connections in each cell to two.
In order to better train the search stage architecture, we ensure that the network \emph{only} optimizes the network parameters in the first 8 epochs. After the output results are relatively stable, the network parameters and architecture weights are optimized simultaneously. After optimizing, we can get the corresponding network architecture for the specific dataset.
Our code is based on PyTorch 1.7, performing on Windows 10 with 32GB RAM and a NVIDIA RTX 8000 GPU.

\subsection{Comparison with State-of-the-art Methods }

In order to verify the performance of the proposed method, we conduct experiments with three different dose levels: 2.5\%, 5\%, and 10\%. Meanwhile, several state-of-the-art LDCT restoration methods such as BM3D~\cite{dabov2007image}, DnCNN~\cite{zhang2017beyond}, MAP-NN~\cite{shan2019competitive}, and RED-CNN~\cite{chen2017low} are compared. All the models expect RED-CNN are implemented according to the original paper and employ the original loss functions. In RED-CNN, we employ the same loss as MANAS. The quantitative results on the whole testing set are listed in Table~\ref{tab:quantitative_methods}. It can be found that the networks searched using our method only achieve middle positions in terms of both PSNR and SSIM for all the dose-levels. However, according to current studies~\cite{yang2018low,wang2018image,ledig2017photo}, both PSNR and SSIM cannot always well judge the image quality. Therefore, we add perceptual loss, termed PL, as a metric to show the performance of image detail preserving. We can see the MANAS has the lowest perceptual loss, which shows our MANAS can preserve more image details. To further illustrate the performance of MANAS and the visual effect of image detail restoration, three slices reconstructed using different methods from 10\%, 5\%, and 2.5\% dose-levels are given in Fig.~\ref{fig:denoising_10to2.5}, respectively. It is noticed that as the dose-level reduces, the artifacts and noise become serious and most details are covered. All the methods can suppress the artifacts and noise to a certain degree. In the 3rd row of Fig.~\ref{fig:denoising_10to2.5}, there are obvious streak artifacts near the femurs in the results of BM3D. In the results of DnCNN and MAP-NN, the details are blurred, especially in the 1st and 2nd rows of Fig.~\ref{fig:denoising_10to2.5}. Some contrast-enhanced vessels in the liver indicated by arrows are smoothed. RED-CNN obtains most close performance to ours, but it still suffers from some perceptible over smoothed effects, which leads to the spatial contrast loss. Overall, the proposed model achieves the best visual effects in both artifact reduction and detail preservation.

\subsection{Comparison of Network Parameters}

To assess the complexity of the searched networks, the amount of parameters (model scale) and floating-point operations per second (FLOPs) are adopted. The results are listed in Table~\ref{tab:parameters}. DnCNN is a simple plain CNN aided by one residual connection. MAP-NN has a parameter-sharing generator, which is a lightweight model, but its discriminator is of large scale. The searched networks using our method have a smaller model scale and FLOPs than RED-CNN. It is reasonable to say that the performance of the network has an approximate positive relation to the model scale, and it is very hard to have a satisfactory result with a lightweight model for every dose level.

\begin{table}[t]
\centering
\begin{tabular}{lcc}
\toprule
Method            & Params(M)   & FLOPs(G)  \\ \midrule
MAP-NN (G+D)~\cite{shan2019competitive}      & 0.06+269.58 & 1.96+4.55 \\
DnCNN~\cite{zhang2017beyond}             & 0.56        & 36.51     \\
RED-CNN~\cite{chen2017low}          & 1.85        & 121.2     \\ \hline
MANAS Dose 2.5\% & 2.96        & 114.01    \\
MANAS Dose 5\%   & 1.66        & 105.51    \\
MANAS Dose 10\%  & 2.79        & 98.02     \\ \bottomrule
\end{tabular}
\caption{The parameters in different network architecture. }
\label{tab:parameters}
\end{table}

\subsection{Model Investigation}

\begin{figure}[h]
\begin{center}
\includegraphics[width=1.0\linewidth]{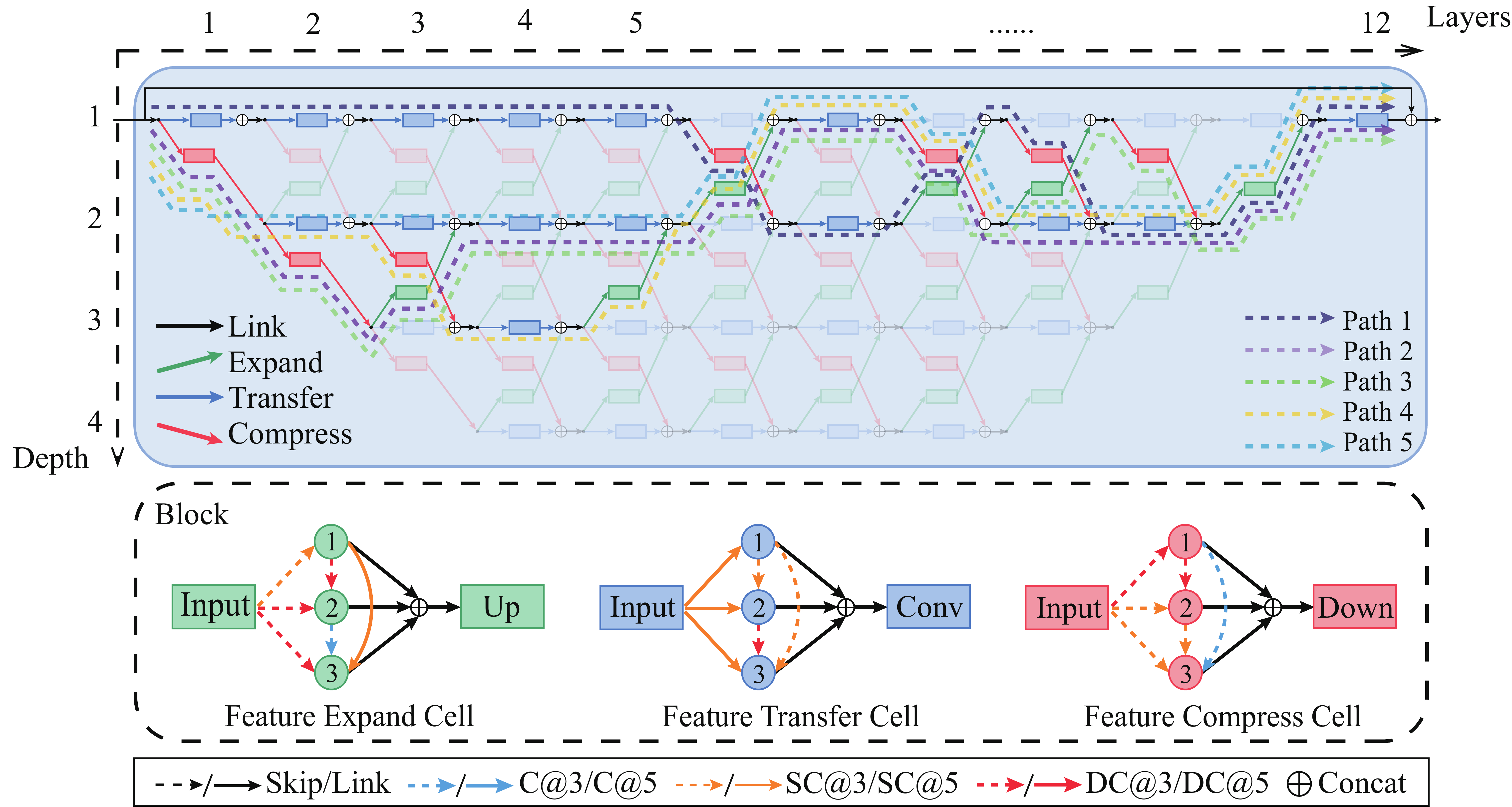}
\end{center}
   \caption{Network architecture result with $L=12$, $D=4$, $K=5$ with 10\% dose data and the corresponding basic blocks.}
\label{fig:network_10}
\end{figure}

\begin{figure}[h]
\begin{center}
\includegraphics[width=1.0\linewidth]{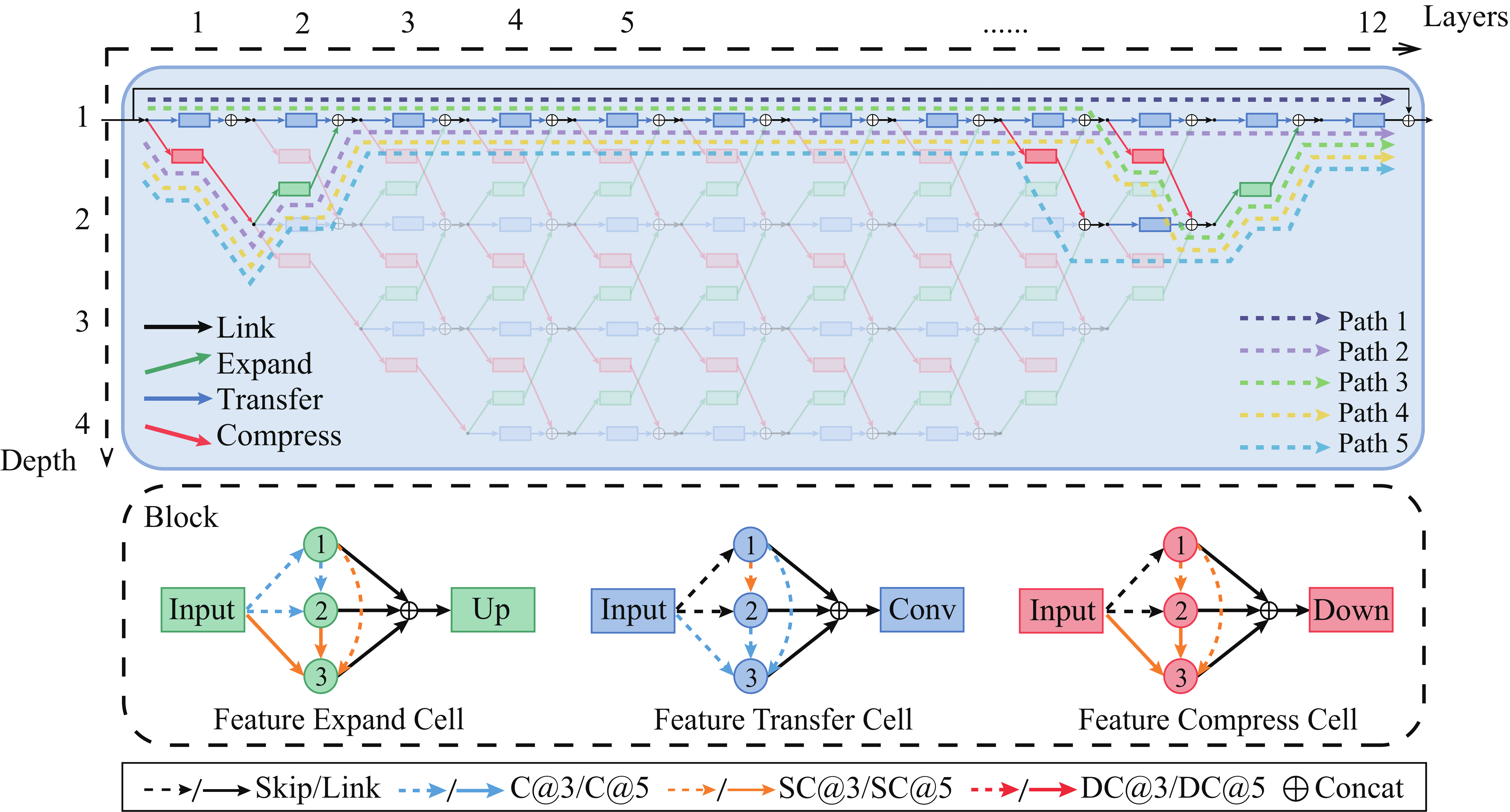}
\end{center}
   \caption{Network architecture result with $L=12$, $D=4$, $K=5$ with 5\% dose data and the corresponding basic blocks.}
\label{fig:network_5}
\end{figure}

\begin{figure}[h]
\begin{center}
\includegraphics[width=1.0\linewidth]{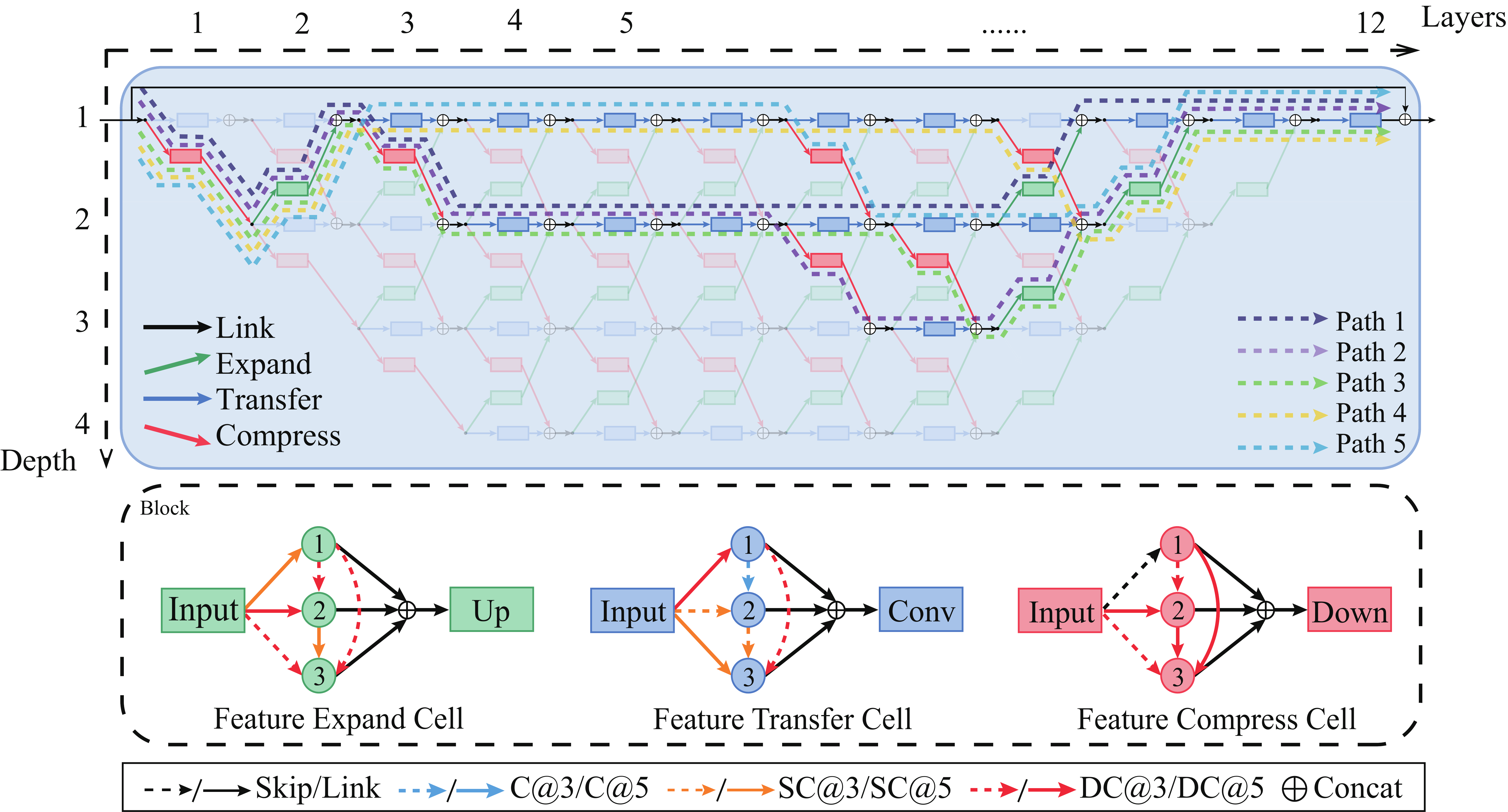}
\end{center}
   \caption{Network architecture result with $L=12$, $D=4$, $K=5$ with 2.5\% dose data and the corresponding basic blocks.}
\label{fig:network_2.5}
\end{figure}

\noindent\textbf{Architecture Analysis}\quad Figs.~\ref{fig:network_10},~\ref{fig:network_5} and~\ref{fig:network_2.5} show the results of sub-network architectures and corresponding cells for different datasets using our method. By looking into the generated network structure, we have the following observations. First, different types of convolutions are included in the final sub-networks, which demonstrates the powerful ability of our model for operation selection. Although separable convolution is efficient in reducing network scale and dilated convolution excels at enlarging the receptive field, normal convolution and skip connection are still selected.
Second, the model collapse, which means the entire cell is full of skip-connection, is well avoided by the early stopping strategy~\cite{liang2019darts+,yan2020neural, zhang2020memory}.
Third, all the searched architectures do not use the fourth depth to build the network, which implies that for the LDCT restoration, simply increasing the network depth cannot always improve the performance.
Last, the network architectures generated for different dose-levels are quite different. The possible reason lies in that the images with different dose-levels are contaminated by artifacts and noise to varying degrees, which may have significant impact on the choice of network architecture.

\begin{figure*}[!htb]
   \begin{center}
   \includegraphics[width=0.9\linewidth]{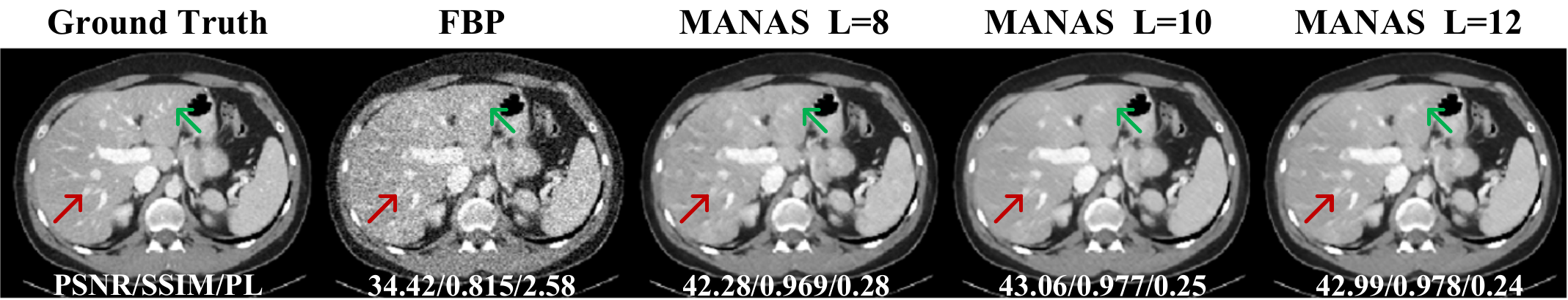}
   \end{center}
      \caption{Image denoising with 5\% dose data using different layer numbers in super-net.}
   \label{fig:denosing_5_layer}
   \end{figure*}

\begin{figure*}[!htb]
   \begin{center}
   \includegraphics[width=0.9\linewidth]{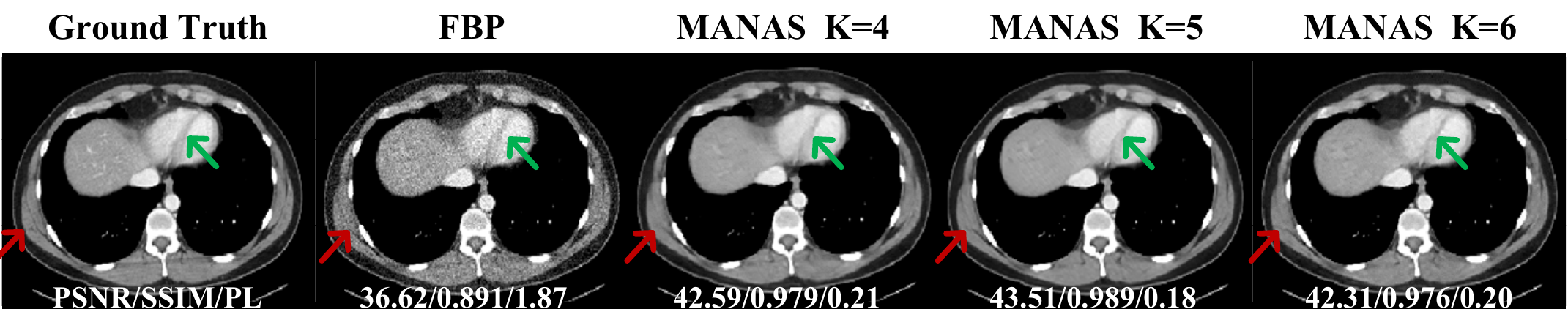}
   \end{center}
      \caption{Image denoising with 5\% dose data using different path numbers in super-net.}
   \label{fig:denoising_5_path}
   \end{figure*}

\begin{table}[htb]
   \centering
   \begin{tabular}{cccc}
   \toprule
   Layer $L$ & PSNR  $\uparrow$         & SSIM $\uparrow$        & PL $\downarrow$ \\ \midrule
   8               & 42.46{\footnotesize $\pm$0.54} & 0.974{\footnotesize $\pm$0.004} & 0.24{\footnotesize $\pm$0.04}\\
   10              & \textbf{44.43{\footnotesize $\pm$0.60}} & \textbf{0.984{\footnotesize $\pm$0.003}} & 0.22{\footnotesize $\pm$0.05}
  \\
   12              & 44.36{\footnotesize $\pm$0.62} & \textbf{0.984{\footnotesize $\pm$0.003}} & \textbf{0.22{\footnotesize $\pm$0.04}} \\ \bottomrule
   \end{tabular}
   \caption{Quantitative result (Mean$\pm$SD) with different numbers of initial layers on the testing set.}
   \label{tab:quantitative_layers}
\end{table}
\noindent\textbf{Effects of Number of Super-network Layers $L$}\quad To evaluate the impact of the number of super-network layers $L$, we initialize the super-network with 8, 10 and 12 layers respectively. The LDCT images with 5\% dose are used as the training and testing sets. The other parameters are fixed as we mentioned before. The statistical quantitative results for the whole testing set are given in Table~\ref{tab:quantitative_layers}. It can be noticed that when the number of layers is greater than 8, the improvement is not quite limited. The qualitative results of one representative slice are shown in Fig.~\ref{fig:denosing_5_layer}. It can be observed that the results using 10 and 12 layers as initial architectures obtain better visual effects than the one with 8 layers with fewer artifacts, which is coherent with the quantitative scores. The arrows indicate some obvious differences. The result with $L=12$ recovers more details than the one with $L=10$. Based on this observation, $L$ in our model is initialized to 12.

\begin{table}[t]
\centering
\begin{tabular}{cccc}
\toprule
Path $K$ & PSNR $\uparrow$           & SSIM $\uparrow$        & PL $\downarrow$ \\ \midrule
4              & 43.24{\footnotesize $\pm$0.54} & 0.976{\footnotesize $\pm$0.004} & 0.27{\footnotesize $\pm$0.05} \\
5              & \textbf{44.36{\footnotesize $\pm$0.62}} & \textbf{0.984{\footnotesize $\pm$0.003}} & \textbf{0.22{\footnotesize $\pm$0.04}} \\
6              & 42.88{\footnotesize $\pm$0.58} & 0.973{\footnotesize $\pm$0.005} & 0.27{\footnotesize $\pm$0.05} \\ \bottomrule
\end{tabular}
\caption{Quantitative result (Mean$\pm$SD) with different $K$ on the testing set.}
\label{tab:quantitative_path}
\end{table}
\noindent\textbf{Effects of the Numbers of Paths $K$}\quad We evaluate the impact of the number of paths $K$ to form the final network. Three different numbers of paths are tested in the experiments, including 4, 5, and 6. The other parameters are fixed and $L=12$. The statistical quantitative results are listed in Table~\ref{tab:quantitative_path}. When $K=5$, the searched network achieves the best scores. One typical thoracic slice processed using networks with different numbers of paths are illustrated in Fig.~\ref{fig:denoising_5_path}. It is easy to notice that the result reconstructed by the network with $K=5$ can better visualize the structural details indicated by the arrows in Fig.~\ref{fig:denoising_5_path}, which is also confirmed by the quantitative metrics. Based these results, it is suggested that the optimal number of paths may vary when given a specific dataset or super-net architecture.


\section{Conclusion}\label{sec:conclusion}

In this paper, we proposed a multi-scale and multi-level gradient based NAS for low-dose CT denoising. The proposed method searches the network architecture in both cell- and network-levels, which provides an extended search space and is more flexible and efficient than traditional NAS methods. Meanwhile, to leverage the multi-scale features, three different feature fusion cells are introduced. The searched networks are evaluated on the datasets with different dose levels and demonstrate better performance in terms of image structural details than several handcrafted state-of-the-art models, which reflects the robustness and effectiveness of MANAS. In addition, different configurations of MANAS further illustrate the influence of different scale features on image detail restoration.

We acknowledge some limitations of our method. First, since we expand the search space, in spite of DARTS, it is time- and resource-consuming for training. Meanwhile, since we need to train the model twice, one for architecture search and the other for parameter optimization, the computational burden is further aggravated. In this study, it takes two GPU days (RTX 8000) to train the model. The other one lies in that the proposed three different cells have the same structures, which may limit the possible results. In the future, designing a more general search space will be the next step.

\clearpage
{\small
\bibliographystyle{ieee_fullname}
\bibliography{manasbib}
}

\end{document}